\documentclass[twocolumn,epjc3]{svjour3}  
\smartqed  
\RequirePackage{graphicx}
\usepackage{xcolor}
\usepackage{booktabs}
\usepackage{amssymb}
\usepackage{amsmath}
\journalname{Eur. Phys. J. C}

\newcommand{\etal}{{\em et al.}}
\newcommand{\eg}{{\em e.g.}}
\newcommand{\ie}{{\em i.e.}}

\begin{document}

\title{Alternative to the application of PDG scale factors}

\author{Jens Erler\thanksref{e1,addr1,addr2} \and Rodolfo Ferro-Hern{\'a}ndez\thanksref{e2,addr1}}
	
\thankstext{e1}{erler@fisica.unam.mx}
\thankstext{e2}{rferrohernandez00@gmail.com}

\institute{Departamento de F\'{\i}sica Te\'{o}rica, 
Instituto de F\'{\i}sica,
Universidad Nacional Aut\'{o}noma de M\'{e}xico, 
04510 CDMX,
M\'{e}xico\label{addr1}
\and
PRISMA$^+$ Cluster of Excellence and Helmholtz Institute Mainz, Johannes Gutenberg-Universit\"at, 55099 Mainz, Germany\label{addr2}}
	
\date{}

\maketitle

\begin{abstract}
The Particle Data Group recommends a set of procedures to be applied when discrepant data are to be combined. 
We introduce an alternative method based on a more general and solid statistical framework, providing a robust way to include possible unknown systematic effects interfering with experimental measurements or their theoretical interpretation.
 The limit of large data sets and practical cases of interest are discussed in detail.


\end{abstract}

\section{Introduction}
\label{intro}
In any field of science, it is often the case that a number of data points or data sets need to be combined in order to achieve a greater overall precision.
Now, data naturally fluctuate and it is not uncommon that one or several data points may appear discrepant or outlying with respect to the bulk of the data.
This is not necessarily a concern, \eg, if the results of the individual measurements or observations are known to be dominated by the statistical uncertainty,
or even in the presence of significant systematic effects, as long as their associated uncertainties can be reliably estimated.
On the other hand, if the observed discrepancies are suspiciously large or plentiful, one may worry that some unknown systematic effect or 
unjustified but hidden assumption might have moved the central value of one or more observations.
In that latter case, a more conservative handling of the data and its combination would be called for.

Of course, it is impossible to know independently which of the aforementioned situations
--- larger than expected random fluctuations, unknown systematic effect(s), or both ---
one is facing, or which of the individual data (sub)sets could be at fault.  
As a remedy, the Particle Data Group\footnote{The PDG collects, evaluates, averages and fits particle physics data world-wide 
and assesses their implications and interpretations in a large number of dedicated reviews.} (PDG)~\cite{PDG2018} proposed a set of rules 
according to which the uncertainty of an average is to be enlarged by a scale factor $S$, while the central values are to remain unchanged by fiat.
Assuming Gaussian errors, in a first step the reduced $\chi^2$ is computed as twice the log-likelihood of the minimum divided by $N_{\rm eff}$, 
where $N_{\rm eff}$ is the effective number of degrees of freedom given by the number of observations (data points), $N$, 
minus the number of independent fit parameters.   
Thus, for the most common case of a simple average of one parameter, $N_{\rm eff} = N -1$:
\begin{enumerate}
\item If the reduced $\chi^2$ is smaller than unity, the results are accepted and there is no scaling of errors.
\item If the reduced $\chi^2$ is larger than unity, and the experiments are of comparable precision, 
then all errors are re-scaled by a common factor $S$, given by the reduced $\chi^2$, \ie, $S = \sqrt{\chi^2/N_{\rm eff}}$. 
\item If some of the individual errors are much smaller than others, then $S$ is computed from only the most precise experiments. 
The criterium for these is given with reference to an {\em ad hoc}\/ cutoff value. 
\end{enumerate}
Given that the rationale for a procedure such as this one, is to err on the conservative side,
one immediate objection is that if there is only one data point then no conservative scaling will be applied,
even though in this case one is most exposed to a potential problem as there is no control measurement. 

Another problem is that the set of individual data points is not well-defined.
In principle, one may combine certain data subsets first, such as from different data taking periods or different decay channels 
obtained by the same experimental apparatus, or combine identical channels obtained by different detectors and average these is a second step.
Conversely, one could split up the available results into more but less precise individual entries.
While this has no impact on ordinary maximum likelihood analyses, 
it will generally dilute or enlarge the reduced $\chi^2$ value on which the $S$ factors are based upon.
In fact, applying PDG scale factors to data points of which some have already undergone the scale factor treatment
(typically, by the experimental collaboration) then this kind of iteration {\em does\/} generally change the central value of the combination.
Also note that the prescription according to which reduced $\chi^2$ values greater and smaller than unity are being treated differently
generates an unnecessary dichotomy.

In this paper we present an alternative which shares some of the features of the PDG recommendation while improving on others.
The framework is a hierarchical model within Bayesian parameter inference~\cite{gelmanbda04}.
The basic idea is that individual data points are not considered independently and identically distributed ({\em iid}\/), 
but rather independently and similarly distributed, in the sense that the parent distributions are permitted to vary to some extent 
to allow for unknown effects that may or may not be different from one data point (measurement) to another.
Thus, we propose a hierarchical model where each measurement is assumed to determine a different parameter, 
each considered as having arisen as a random draw from a common parent distribution described in turn in terms of {\em hyper-parameters}.

A similar approach is widely used in the biological sciences when estimating treatment effects by combining several studies performed under similar 
but not identical conditions~\cite{Tarone,DempsterSelwynandWeeks}, in what is often referred to 
as {\em meta-analysis}~\cite{StatMethods,StudyMetanalysisCancer,rarediseases}.
In these cases the experimental conditions can vary slightly, so that the individual studies may be affected by different unknown biases.

Several authors within the physics community introduced attempts to incorporate the effects of unknown error sources when combining data.
For example, Ref.~\cite{Cowan:2018lhq} finds results similar to the ones in our work, but within a frequentist approach. 
Ref.~\cite{DAgostini:1999niu} models the probability of underestimating the experimental error by including a different scale factor for each measurement, 
which is in turn randomly drawn from a prior distribution.
Very recently it was shown~\cite{MukhopadhyaySubhadeepandFletcherDouglas} that it is even possible to test the shape of the prior distribution, 
and not just to constrain the values of its parameters.  
We leave this kind of more complete analysis for the future.  

In the next section we summarize the formalism of Bayesian hierarchical modeling using the notation of Ref.~\cite{gelmanbda04}.
The rest of the paper introduces our approach, illustrated by a number of examples and reference cases.

\section{Bayesian Inference}
\label{sec:1}
\subsection{The non-hierarchical model}
Suppose that we want to determine a parameter $\theta$ from an experimental measurement or observation, and to be specific, 
that the likelihood for the outcome $y$ of such an experiment can be described as a Gaussian with central value $\theta$ and standard deviation~$\sigma$,
\begin{equation}
p(y|\theta,\sigma) = \mathcal{N}(y|\theta,\sigma),
\end{equation} 
where,
\begin{equation}
\mathcal{N}(y|\theta,\sigma) \equiv \frac{1}{\sqrt{2\pi}\sigma}e^{-\frac{1}{2\sigma^2}(y-\theta)^2}\ .
\end{equation} 
The posterior distribution for the parameter $\theta$ can be obtained through Bayes' theorem,
\begin{equation}
p(\theta|y,\sigma)\propto p(y|\theta,\sigma)p(\theta),
\end{equation}
where $p(\theta)$ is the {\em prior probability distribution\/} of $\theta$.
It is very convenient to assume $p(\theta)$ to be a {\em conjugate prior}, 
which means that the posterior distribution will fall within the same family of functions as the prior.
Thus, in our case we adopt the prior,
\begin{equation}
\theta \sim \mathcal{N}(\tilde\mu,\tilde\tau),
\end{equation}
yielding the posterior,
\begin{equation}
\label{eq:normal}
p(\theta|y,\sigma,\tilde\mu,\tilde\tau) = \frac{1}{\sqrt{2\pi}\sigma_{\tilde\tau}}e^{-\frac{1}{2\sigma^2_{\tilde\tau}}(\theta-\theta_{\tilde\tau})^2},
\end{equation}
where,
\begin{equation}
\frac{1}{\sigma^2_{\tilde\tau}}\equiv\frac{1}{\sigma^2}+\frac{1}{\tilde\tau^2}\ ,
\end{equation}
is the sum of precisions of the prior and the experimental result, while
\begin{equation}
\theta_{\tilde\tau}\equiv\left(\frac{1}{\sigma^2}+\frac{1}{\tilde\tau^2}\right)^{-1}\left(\frac{y}{\sigma^2}+\frac{\tilde\mu}{\tilde\tau^2}\right),
\end{equation} 
is the precision averaged central value. 
Clearly, if the experiment has a small error, $\sigma \ll \tilde\tau$, it will dominate~$\theta_{\tilde\tau}$.   
In the limit $\tilde\tau \to \infty$, the prior is called {\em non-informative}.

\begin{figure}
\begin{centering}
\includegraphics[width=0.3\textwidth]{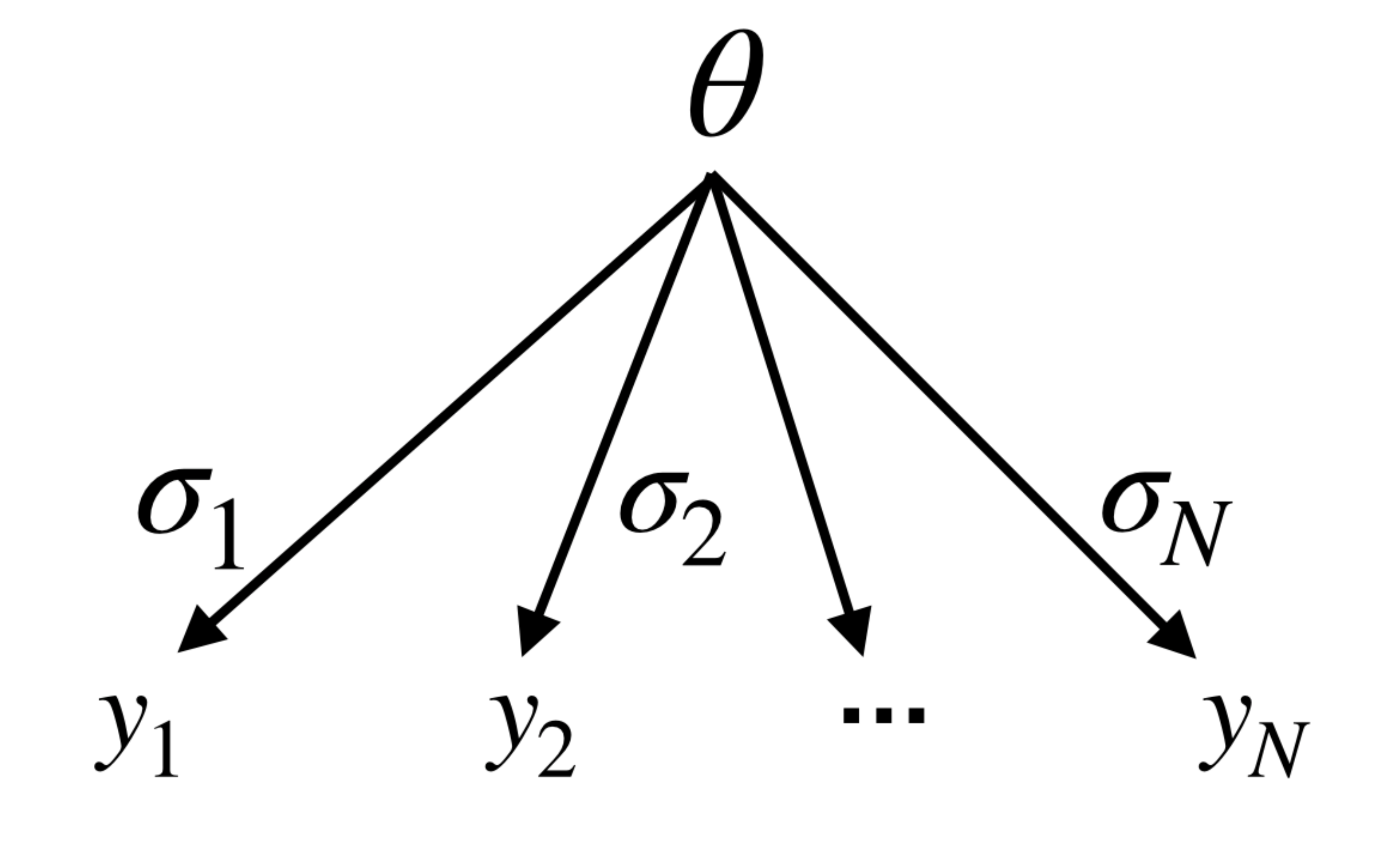}
\caption{Ordinary averaging. 
We assume that the $y_i$ are random outcomes of measurements of the {\em same\/} parameter~$\theta$.}
\label{fig:completepoolingdiagram}   
\end{centering}
\end{figure}

Now, let us include further such experiments with central values $y_i$ and total errors $\sigma_i$,
all measuring the same quantity~$\theta$, as illustrated in Fig.~\ref{fig:completepoolingdiagram}.
For simplicity, we assume that the $\sigma_i$ are mutually uncorrelated.
The posterior distribution $p(\theta|y_i,\sigma_i,\tilde\mu,\tilde\tau)$ is again given by Eq.~(\ref{eq:normal}), but now with
\begin{equation}
\frac{1}{\sigma^2_{\tilde\tau}} = \sum^{N}_{i=1} \frac{1}{\sigma^2_{i}} + \frac{1}{\tilde\tau^2}\ ,
\end{equation}
and 
\begin{equation}
\theta_{\tilde\tau} = \sigma^2_{\tilde\tau} \left(\sum^{N}_{i=1}\frac{y_i}{\sigma^2_{i}}+\frac{\tilde\mu}{\tilde\tau^2}\right).
\end{equation}
Obviously, the uncertainty $\sigma_{\tilde\tau}$ in $\theta$ decreases strictly monotonically with the inclusion of more experiments. 
Nevertheless, if one or several of the experiments was subject to a number of systematic effects that was neither corrected for,
nor accounted for in the individual uncertainties $\sigma_i$, then the experiments are (effectively) not measuring the same quantity, 
and $\sigma_{\tilde\tau}$ would be underestimated.
In other words, each experiment can be viewed as measuring different parameters $\theta_i$,
which are, however, not entirely independent of each other, since after all, the experiments were supposed to constrain the same $\theta$. 
We will now review hierarchical Bayesian modeling, and propose it as a systematic method to interpolate 
between the extreme and rarely realistic cases of all $\theta_i$ being either equal or else entirely independent of each other.

\subsection{The hierarchical model}
\label{sec:2}
\begin{figure}[t]
\begin{centering}
\includegraphics[width=0.3\textwidth]{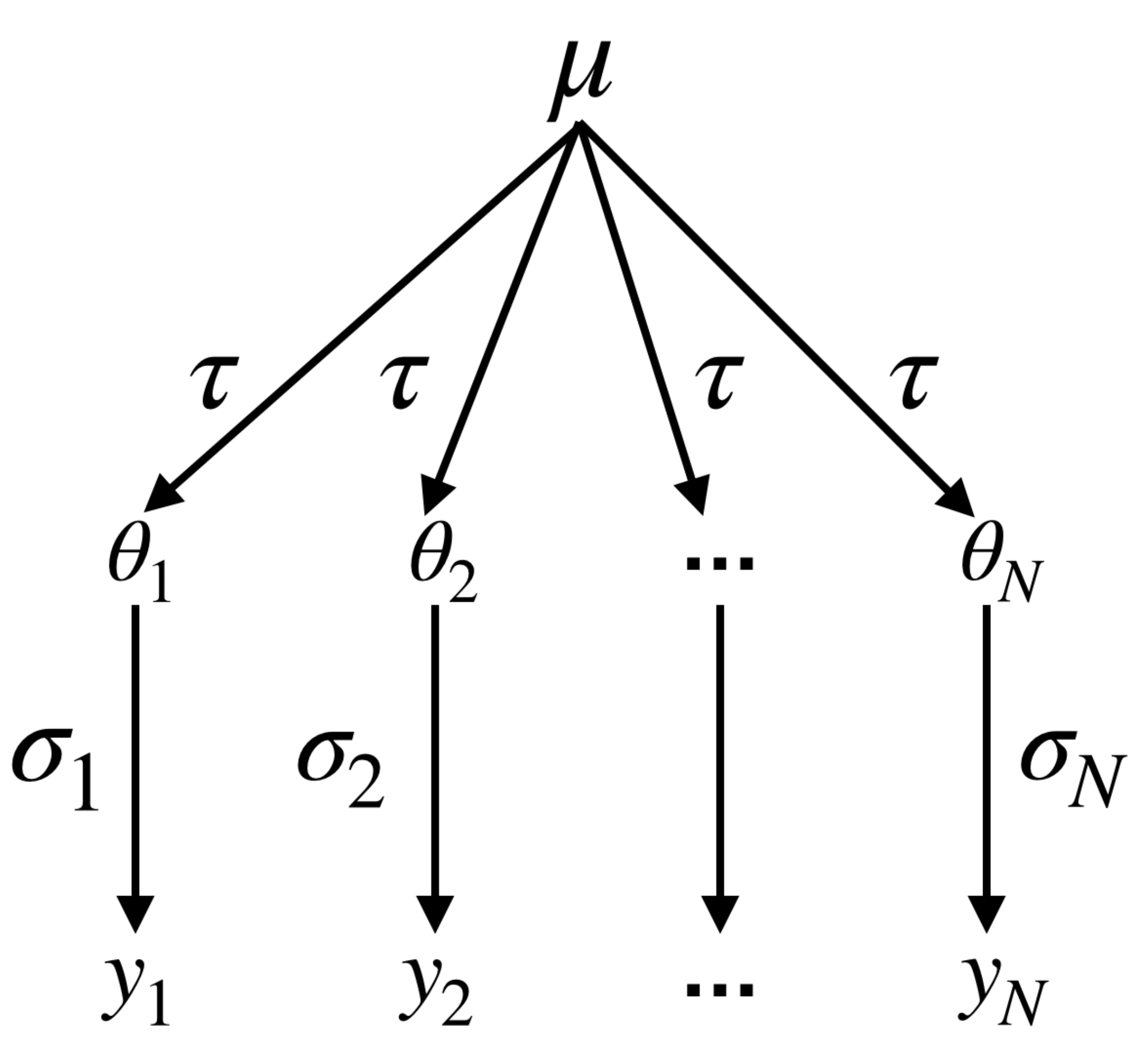}
\caption{Hierarchical model. 
Each experimental parameter $\theta_i$ arises from a random draw from a parent distribution with hyper-parameters $\mu$ and $\tau$, and each experimental 
central value $y_i$ is then considered to be the result of a random draw from a Gaussian distribution with central value $\theta_i$ and error~$\sigma_i$.}
\label{fig:hierarchichalmodeldiagram} 
\end{centering}
\end{figure}

This is achieved by considering each $\theta_i$ to be the result of a random draw from a parent distribution,
\begin{equation}
\label{eq:hyperdraw}
p(\theta_i)=\int p(\theta_i|\mu,\tau)p(\mu,\tau)d\mu d\tau,
\end{equation}
where $p(\mu,\tau)$ is the {\em hyper-prior distribution\/} for what are now called the hyper-parameters $\mu$ and $\tau$.
We sketch this model in Fig.~\ref{fig:hierarchichalmodeldiagram}.  
Note that Eq.~(\ref{eq:hyperdraw}) implies the property of ex-changeability between the $\theta_i$, {\em i.e.\/} symmetry under $\theta_i\leftrightarrow\theta_j$. 
From Bayes' theorem one has,
\begin{equation}
p(\theta_i,\mu,\tau|y_i,\sigma_i)\propto p(y_i|\theta_i,\sigma_i)p(\theta_i|\mu,\tau)p(\mu,\tau),
\end{equation}
and explicitly in the Gaussian case,
\begin{equation}
p(\theta_i,\mu,\tau|y_i,\sigma_i)\propto \prod_{i=1}^{N}\mathcal{N}(y_i|\theta_i,\sigma_i)\mathcal{N}(\theta_i|\mu,\tau) p\left(\mu,\tau\right).
\end{equation}
Marginalizing over $\theta_i$ one finds the ``master" equation,
\begin{equation}
p(\mu,\tau|y_i,\sigma_i)\propto\prod_{i=1}^{N}\mathcal{N}(\mu|y_i,\sigma^2_i+\tau^2)p(\mu,\tau).
\label{equationtotal}
\end{equation}
We will use it to compute the posterior distribution of the hyper-parameters, once a hyper-prior is chosen. 
For example, assuming a flat prior for $\mu$ and $\tau$, we can integrate over $\mu$ to find,
\begin{equation}
p(\tau|y_i)\propto\left(\sum^{N}_{i=1}\frac{1}{\sigma^2_{i}+\tau^2}\right)^{-\frac{1}{2}}\prod_{i=1}^{N}\mathcal{N}(\hat{\mu}|y_i,\sigma^2_i+\tau^2)
\label{equationtau},
\end{equation} 
where,
\begin{equation}
\hat{\mu}=\left(\sum^{N}_{i=1}\frac{1}{\sigma^2_{i}+\tau^2}\right)^{-1}\sum^{N}_{i=1}\frac{y_i}{\sigma_i^2+\tau^2}\ .
\end{equation}
The parameter $\tau$ quantifies general differences in the $\theta_i$. 
If $\tau=0$, the experiments measure the same parameter, \ie, $\theta_i=\theta_j$. 
For $\tau\rightarrow\infty$, each one measures a completely independent parameter $\theta_i$.

From the master equation one can see that the parameter of interest is $\mu$. 
If $\tau=0$ the posterior distribution for $\mu$ reduces to the ordinary likelihood for parameter estimation given in Eq.~(\ref{eq:normal}) with $\tilde\tau \to \infty$.
The full posterior distribution for $\mu$ can be obtained integrating Eq.~(\ref{equationtotal}) numerically over $\tau$.  
If there are large unknown systematic effects, then the most likely values of $\tau$ will differ from zero,
which leads to the important result of increasing the error in $\mu$.

\subsection{The hyper-prior}
We propose a hyper-prior which is $\mu$-independent, \ie, $p(\mu,\tau) = p(\tau)$, 
and that interpolates smoothly between a flat and a sharply peaked $\tau$ distribution,
\begin{equation}
\label{hyperprior}
p(\tau) d\tau^2\propto\prod_{i=1}^{N}\left[\frac{1}{\sigma^2_i+\tau^2}\right]^\frac{\alpha}{2N} d\tau^2.
\end{equation}
This form will prove to be useful due to the simple interpretation of $\alpha$ in terms of the number of degrees of freedom,
and the possibility to obtain closed analytical formulas for the posterior distribution of $\mu$.  
We remark that in Bayesian methods one needs to specify a prior that cannot be determined from first principles.
Here we have chosen a prior with a simple analytical form interpolating between a flat prior and $\tau = 0$. 
Very interestingly, while this prior is only one of many possible choices, it turns out that it coincides with Jeffrey{'}s prior in a certain limit.  
We will return to this at the end of Section~\ref{othermodels}.

It is interesting to study the effect of this kind of prior on the tails of the posterior density of $\mu$.
Integrating Eq.~(\ref{equationtotal}) over $\tau$ produces the posterior density of $\mu $ given the data,
\begin{equation}
p(\mu|y_i)\propto\int_{0}^{\infty}\prod_{i=1}^{N}\left(\sigma^2_i+\tau^2\right)^{-\frac{1}{2}(1+\frac{\alpha}{N})}e^{-\frac{(\mu-y_i)^2}{2(\sigma^2_i+\tau^2)}}d\tau^2.
\end{equation}
For large $\mu$, the exponential suppression factor favors large values of $\tau$, so that,
\begin{equation}
p(\mu|y_i)\sim\int_{0}^{\infty} \tau^{-(N+\alpha)}e^{-\frac{N\mu^2}{2\tau^2}}d\tau^2,
\end{equation}
and after a change of variables $u^2\equiv\mu^2/\tau^2$, 
\begin{equation}
p(\mu|y_i)\sim \mu^{-(N+\alpha-2)}.
\end{equation}
We observe that the usual exponential suppression of $\mu$ in the tails has turned into a milder power law suppression 
which increases with the effective number of degrees of freedom, \ie, in our case the number or measurements, $\nu\equiv N+\alpha-2$.

\begin{figure}
	\begin{centering}
		\includegraphics[width=0.5\textwidth]{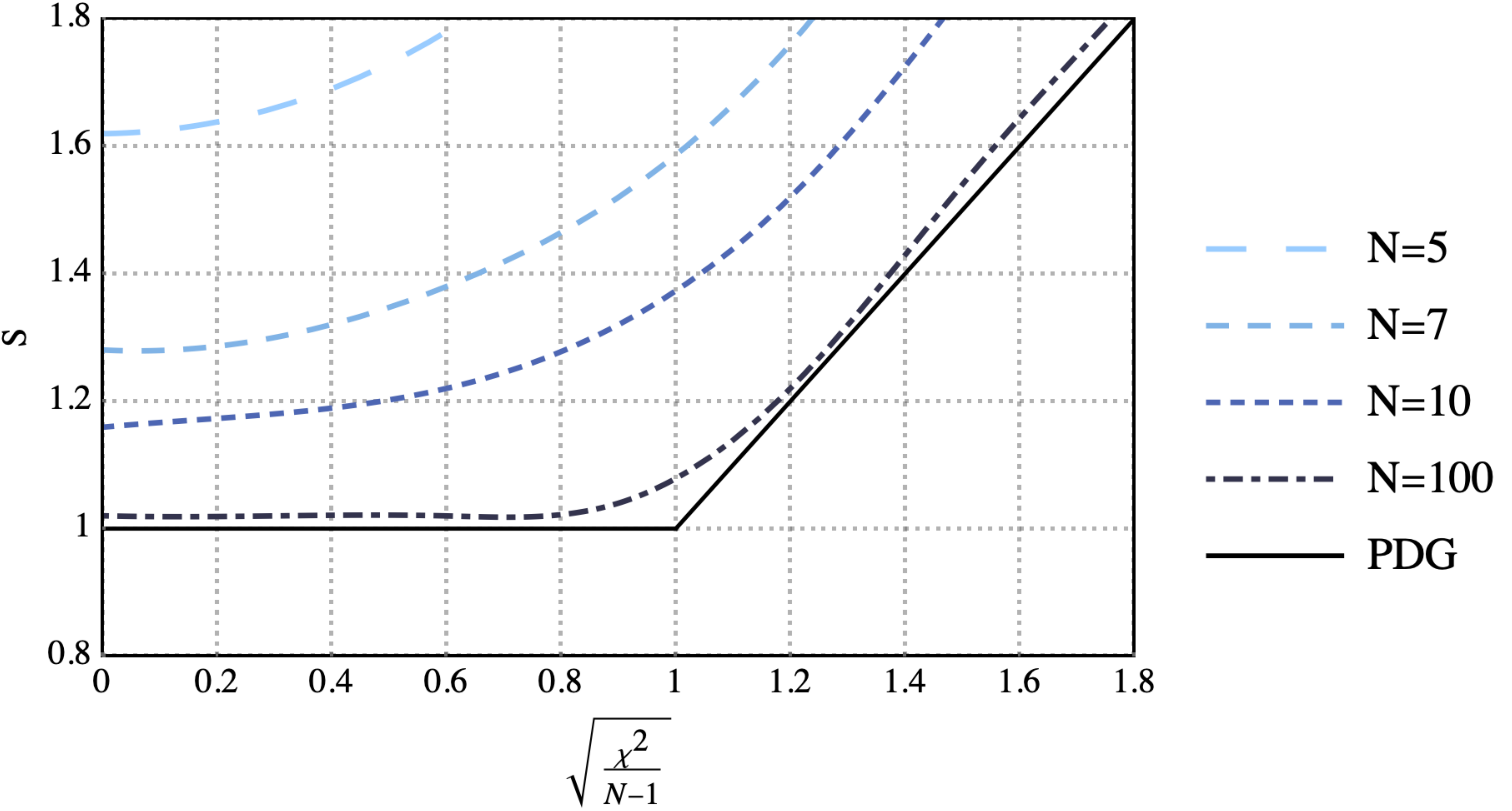}
		\caption{Scale factor versus the square root of the reduced $\chi^2$. We employed $\alpha = 0$.}
		\label{fig:scaleequal}
	\end{centering}
\end{figure}

\section{Experiments with errors of the same size}
When all errors are equal, $\sigma_i = \sigma_j\equiv \sigma$, we obtain an analytical formula which illustrates how the PDG scale factor
re-emerges for large data sets.
The master equation reads in this case,
$$
p(\mu,\tau|y_i)\propto\left(\sigma^2+\tau^2\right)^{-\frac{\nu+2}{2}} \exp\left[ -\frac{\sum_{i=1}^{N}(\bar{y}_i-\mu)^2}{2(\sigma^2+\tau^2)} \right],
$$
or simply,
\begin{equation}
p(\mu|y_i)\propto\intop^{\infty}_{0}\left(\sigma^2+\tau^2\right)^{-\frac{\nu+2}{2}} \exp\left[ -\frac{\sigma^2 \chi^2}{2(\sigma^2+\tau^2)} \right] d\tau^2 ,
\end{equation}
where we defined,
\begin{equation}
\chi^2\equiv\chi^2(\mu)\equiv \sum_{i=1}^N\frac{(\mu-\bar{y}_i)^2}{\sigma^2}\ ,
\end{equation} 
which is the usual $\chi^2$ function. 
Changing variables, 
\begin{equation}
u\equiv\frac{\sigma^2 \chi^2(\mu)}{2(\tau^2+\sigma^2)}\ ,
\end{equation}
we obtain,
\begin{equation}
p(\mu|y_i)\propto(\chi^2)^{-\frac{\nu}{2}}\intop^{\chi^2/2}_{0}u^{\frac{\nu}{2}-1} e^{-u}du \propto
(\chi^2)^{-\frac{\nu}{2}} F^{\nu}(\chi^2),
\label{postmu}
\end{equation}
which is the master formula in this case in terms of the cumulative distribution function $F$ for a $\chi^2$ distribution with $\nu$ degrees of freedom. 
This equation implies an interesting result.  
Since $p(\mu|y_i)$ depends on $\mu$ only through $\chi^2(\mu)$, we have
\begin{equation}
\frac{dp(\mu|y_i)}{d\mu}=\frac{dp(\mu|y_i)}{d\chi^2}\frac{d\chi^2}{d\mu}\ ,
\end{equation}
so that the mode of the distribution is the same as in the usual case, \ie, at the value of $\mu$ where $\chi^\prime(\mu)^2=0$. 
Thus, 
\begin{quote}
For $\sigma_i = \sigma_j$ the posterior distributions of the hierarchical and non-hierarchical models peak at the same location.
\end{quote} 

\begin{figure}
\begin{centering}
\includegraphics[width=0.5\textwidth]{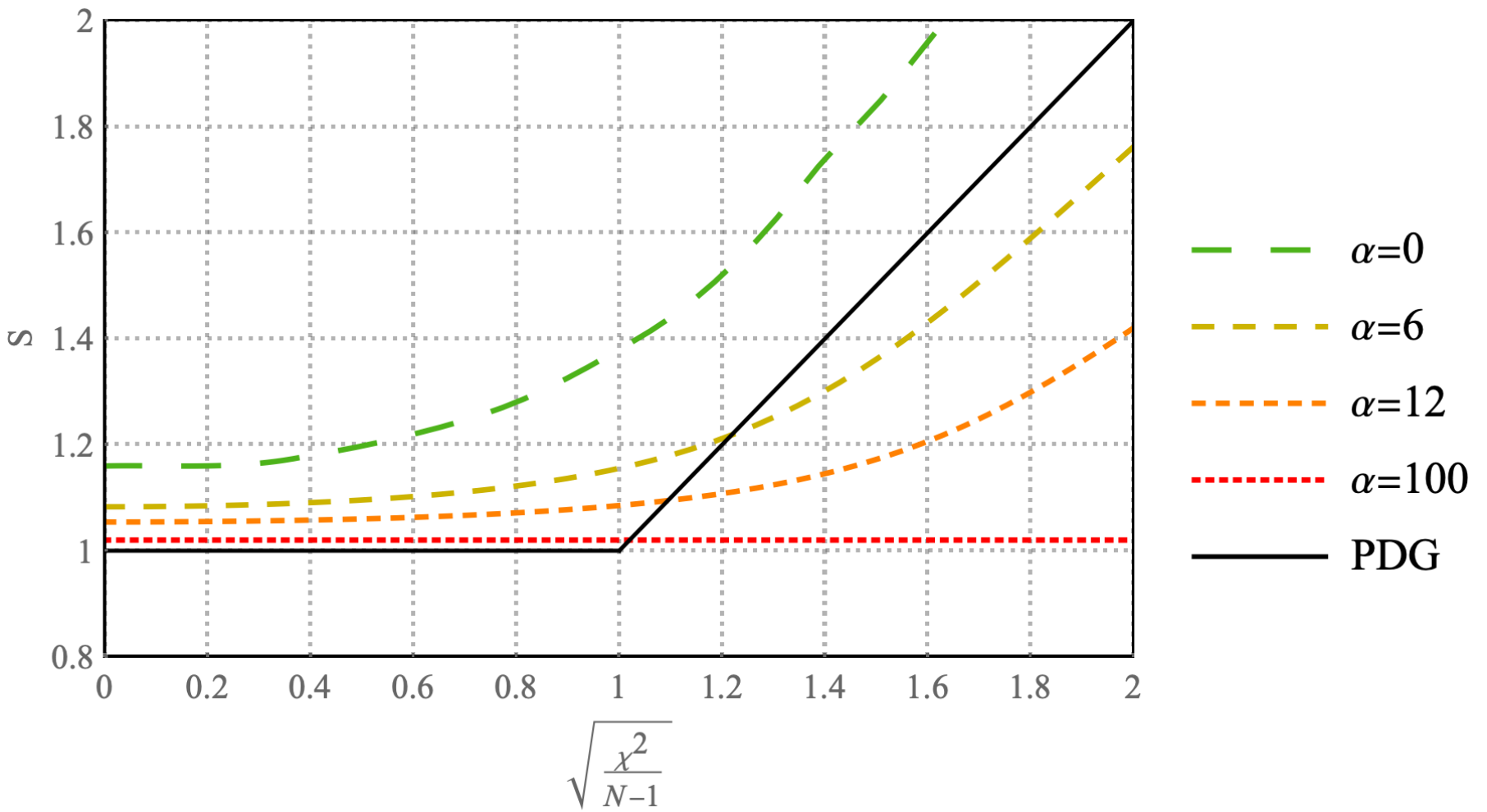}
\caption{Scale factor versus the square root of the reduced $\chi^2$ for the case $N =10$.}
\label{fig:scaleequalalpha}
\end{centering}
\end{figure}

From Eq.~(\ref{postmu}), we can also obtain the scale factor, 
which we define here as the ratio of the sizes of the 68\% highest confidence intervals of the hierarchical and non-hierarchical models.
In Figs.~\ref{fig:scaleequal} and \ref{fig:scaleequalalpha}, we show the scale factor for several values of $\alpha$ and $N$,
from which one can see the similarity to the PDG scale factor for large $N$.
We now turn to the case of a large number of degrees of freedom and the Gaussian approximation.

\subsection{Large number of degrees of freedom}
We rewrite Eq.~(\ref{postmu}) by another change of variables,
\begin{equation}
\frac{\chi^2r}{2}=u,
\end{equation}
so that 
\begin{equation}
p(\mu|y_i)\propto \intop^{1}_{0} \exp\left[-\frac{\nu-2}{2}\left(r\chi^2_{\nu-1}-\ln r\right)\right]dr, 
\label{asympt}
\end{equation}
where we defined $\chi^2_{\nu-1}\equiv \chi^2/(\nu-2)$. 
Thus, large values of $\nu$ suppress the integrand exponentially. 
Depending on the value $r_0=(\chi^2_{\nu-1})^{-1}$ where $ r\chi^2_{\nu-1}-\ln r$ has a minimum, we have two cases:

(1) For $r_0 > 1$ the minimum falls outside the integration limits, and the integral can be approximated by considering values of $r$ near 1, which gives
\begin{equation}
p(\mu|y_i)\propto \frac{e^{-\chi^2/2}}{1-\chi^2_{\nu-1}}\left[1-e^{-\frac{\nu-2}{2}\left(1-\chi^2_{\nu-1}\right)}\right] \sim e^{-\chi^2/2} ,
\end{equation}
We recognize this is the usual likelihood for parameter inference without scaling.
Thus,
\begin{quote}
for $\sigma_i \approx \sigma_j$, $\nu \rightarrow \infty$ and $\chi^2_{\nu-1}(\mu_0)<1$, the hierarchical model implies no scaling of the errors.
\end{quote} 

(2) For $r_0 < 1$ the minimum resides inside the integration region, and the integral can be approximated by considering values of $r$ near $r_0$.
After some algebra,
\begin{equation}
p(\mu|y_i)\propto\left[1+\frac{2}{\nu-1}\frac{(\mu-\mu_0)^2}{2\left(\frac{\sigma^2\chi^2_{\nu}(\mu_0)}{N}\right)}\right]^{-\frac{\nu}{2}},
\end{equation}
which is proportional to the Student-t distribution for $\nu-1$ degrees of freedom, and for very large $\nu$ it can be further approximated by a Gaussian,
\begin{equation}
p(\mu|y_i)= t_{\nu-1}\left(\mu_0,\frac{\sigma^2\chi^2_{\nu}}{N}\right) \sim 
\mathcal{N}\left(\mu_0,\frac{\sigma^2\chi^2_{\nu}}{N}\right).
\end{equation}
This yields another important result,
\begin{quote}
for $\sigma_i\approx\sigma_j$, $\nu \rightarrow \infty$  and $\chi^2_{\nu-1}(\mu_0)>1$, 
the hierarchical model implies a re-scaling of the overall error by $\sigma \rightarrow \sigma\sqrt{\chi^2_{\nu}(\mu_0)}$.
\end{quote} 
It is amusing to note that for large~$\nu$ we recovered the PDG scale factor prescription.
On the other hand, for low values of $\nu$ our model implies larger scalings than recommended by the PDG.
In the next subsection we approximate the distribution of $\mu$ as a Gaussian,
so as to obtain an analytical formula for the scale factor in terms of $\nu$ and the value of $\chi^2$. 

\begin{figure}[t]
\begin{centering}
\includegraphics[width=0.4\textwidth]{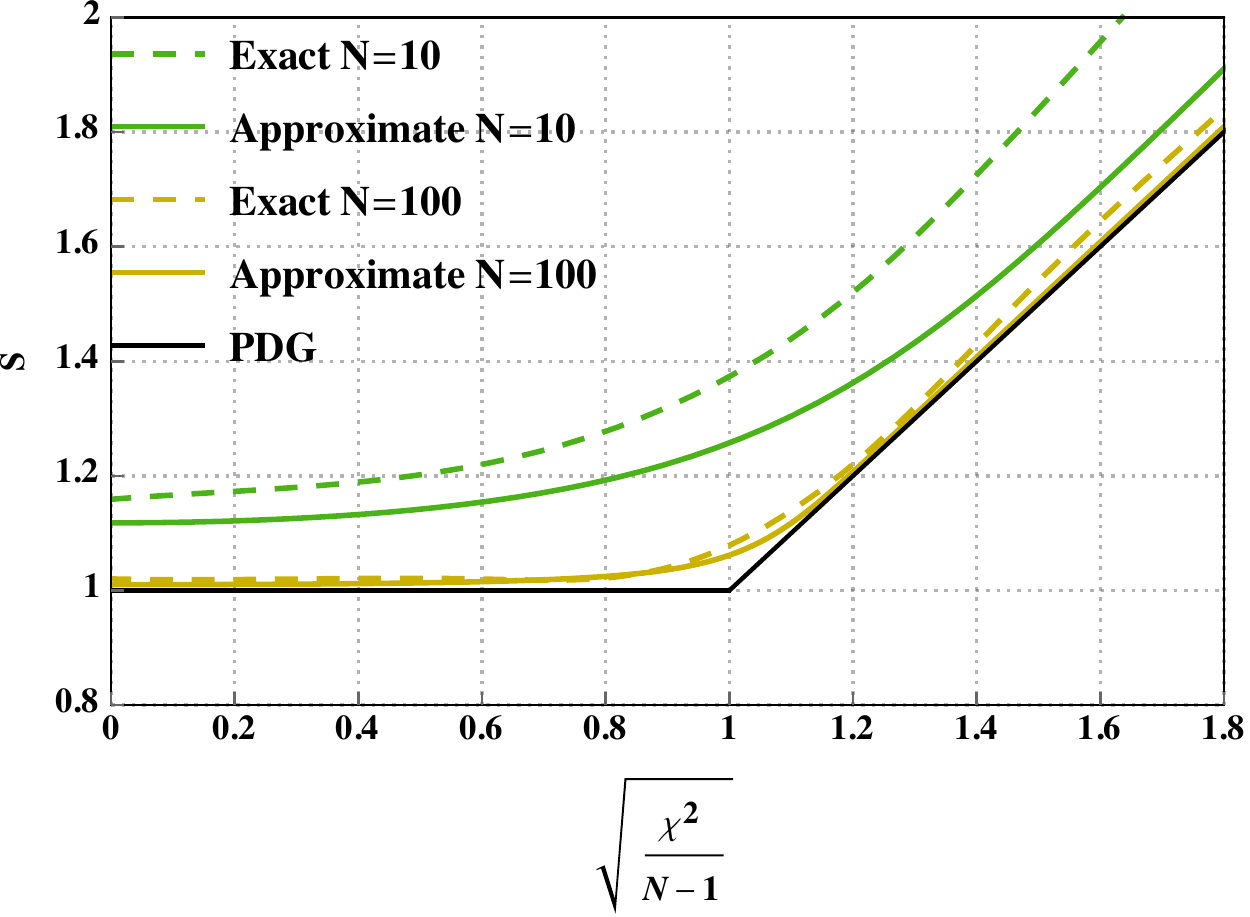}
\caption{Comparison of the exact result with the approximate formula for $\alpha=0$.}
\label{fig:exactapprx}
\end{centering}
\end{figure}

\subsection{Gaussian approximation}
To do so, we expand the logarithm of the posterior distribution $p = p(\mu|y_i)$ in powers of $\mu$ around $\mu_0$,
$$
\ln p=C+\left.\frac{d\ln p}{d\mu}\right|_{\mu_0}(\mu-\mu_0)+\left.\frac{d^2\ln p}{d\mu^2}\right|_{\mu_0}\frac{(\mu-\mu_0)^2}{2}+\cdots
$$
The second term on the right hand side is zero because we are expanding around the maximum. 
The third term can be compared to the corresponding term of the expansion of a Gaussian distribution, which gives
\begin{equation}
\frac{1}{\sigma_{\rm Bayes}^2}\approx-\left.\frac{d^2\ln p}{d\mu^2}\right|_{\mu_0}=-\frac{2N}{\sigma^2}\left.\frac{d\ln p}{d\chi^2}\right|_{\chi^2_0}.
\end{equation}
Using Eq.~(\ref{postmu}) we have,
\begin{equation}
-2 \left.\frac{d\ln p}{d\chi^2}\right|_{\chi^2_0}=\frac{\nu}{\chi^2}-\frac{\left(\chi^2/2\right)^{\left(\frac{\nu}{2}-1\right)} e^{-\chi^2/2}}{\gamma\left(\nu/2,\chi^2/2 \right)}\ ,
\end{equation}
where $\gamma$ is the incomplete Gamma function, defined by
\begin{equation}
\gamma(s,x)\equiv\intop^{x}_{0}t^{s-1}e^{-t}dt.
\end{equation}
As we mentioned before, the scale factor $S_{\rm Bayes}$ is defined as 
the ratio of the sizes of the 68\% highest confidence intervals of the hierarchical and non-hierarchical models.
In the Gaussian approximation we find, 
\begin{equation}
S_{\rm Bayes} \approx\sqrt{N}\, \frac{\sigma_{\rm Bayes}}{\sigma} \approx \sqrt{\frac{\chi^2}{\nu}} 
\left[1+\frac{1}{\sum^{\infty}_{k=1}\frac{(\chi^2)^k \nu!!}{(\nu+2k)!!}}\right]^{\frac{1}{2}},
\end{equation}
where we have used the power series expansion of the incomplete Gamma function,
\begin{equation}
\gamma(s,x)=x^s \mathrm{\Gamma}(s)e^{-x}\sum^{\infty}_{k=0}\frac{x^k}{\mathrm{\Gamma}(s+k+1)}\ .
\end{equation}
In Fig.~\ref{fig:exactapprx} we compare the approximate formula with the exact result. 
As expected, the approximation improves for larger values of $\nu$. 
We are now ready to discuss the general case of unequal errors, $\sigma_i \neq \sigma_j$.

\begin{figure}[t]
\begin{centering}
\includegraphics[width=0.4\textwidth]{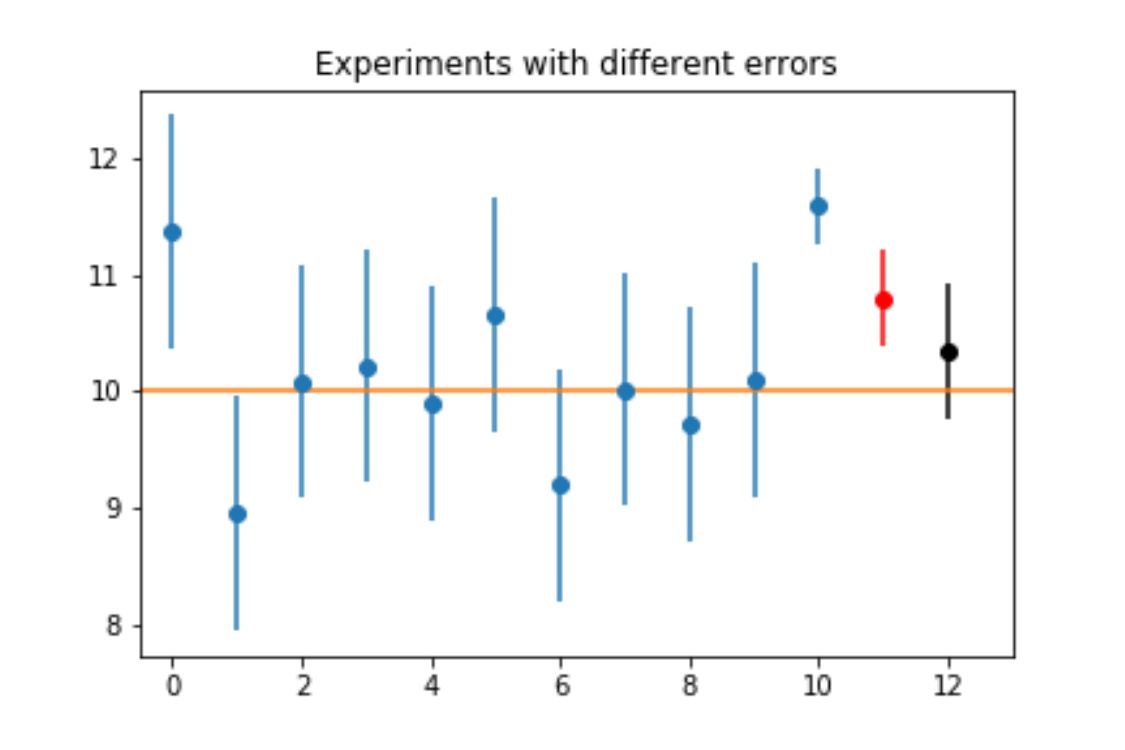}
\caption{The blue points with identical errors originate from a Gaussian distribution centered at 10. 
The last blue point has the same precision as the combination of the previous 10 points, but deviates by about 5 $\sigma$. 
The red point is the ordinary weighted average after PDG scaling.
The black point is obtained using our Bayesian method.}
\label{fig:DifferentErros}
\end{centering}
\end{figure}

\section{Experiments with unequal precisions}
To understand this case, we fix the value of $\tau$ in Eq.~(\ref{equationtotal}).
The distribution of $\mu$ is then Gaussian, with total error,
\begin{equation}
\frac{1}{\sigma^2_t}=\sum^N_{i=1}\frac{1}{\sigma^2_i+\tau^2}\ ,
\end{equation}
and central value,
\begin{equation} 
\mu_0=\left(\sum^N_{i=1}\frac{1}{\sigma^2_i+\tau^2}\right)^{-1}\sum^N_{i=1}\frac{y_i}{\sigma_i^2+\tau^2}\ .
\end{equation}
Thus, experiments with smaller errors are more sensitive to $\tau$ than less precise ones. 
Suppose that $M$ of the experiments have an error $\sigma_M$, and that $\sigma_M$ is much smaller than the error $\sigma$ of the rest of the experiments.
Then, for $\sigma_M\simeq\tau\ll \sigma$ the scaling will mainly  affect the experiments with small errors. Since we were unable to find an analytical formula for the peak or mean of $\tau$, we proceed with a numerical analysis.

As a first example, we randomly generated eleven fictitious measurement points from a Gaussian with standard deviation $\sigma=1$ centered at the value of 10. 
The last point is from a Gaussian centered at $10+5/\sqrt{10}$ with $\sigma_{M}=1/\sqrt{10}$,
which is chosen so that its precision is the same as the combined precision of the other ten.   
The results are shown in Fig.~\ref{fig:DifferentErros}. 
The red point denotes the ordinary weighted average with PDG scaling applied, 
and is pulled away from the horizontal line as a result of the deviating 11th measurement.
The black point, on the other hand, is the average obtained as the result of our Bayesian hierarchical model 
(here we use $\alpha = 10$ to specify our prior).
It is closer to the bulk of data than to the measurement with the smaller error. 
This is a reasonable property, since it is less likely that all the measurements in the bulk had a systematic error in the same direction. 

\begin{figure}[t]
\begin{centering}
\includegraphics[width=0.4\textwidth]{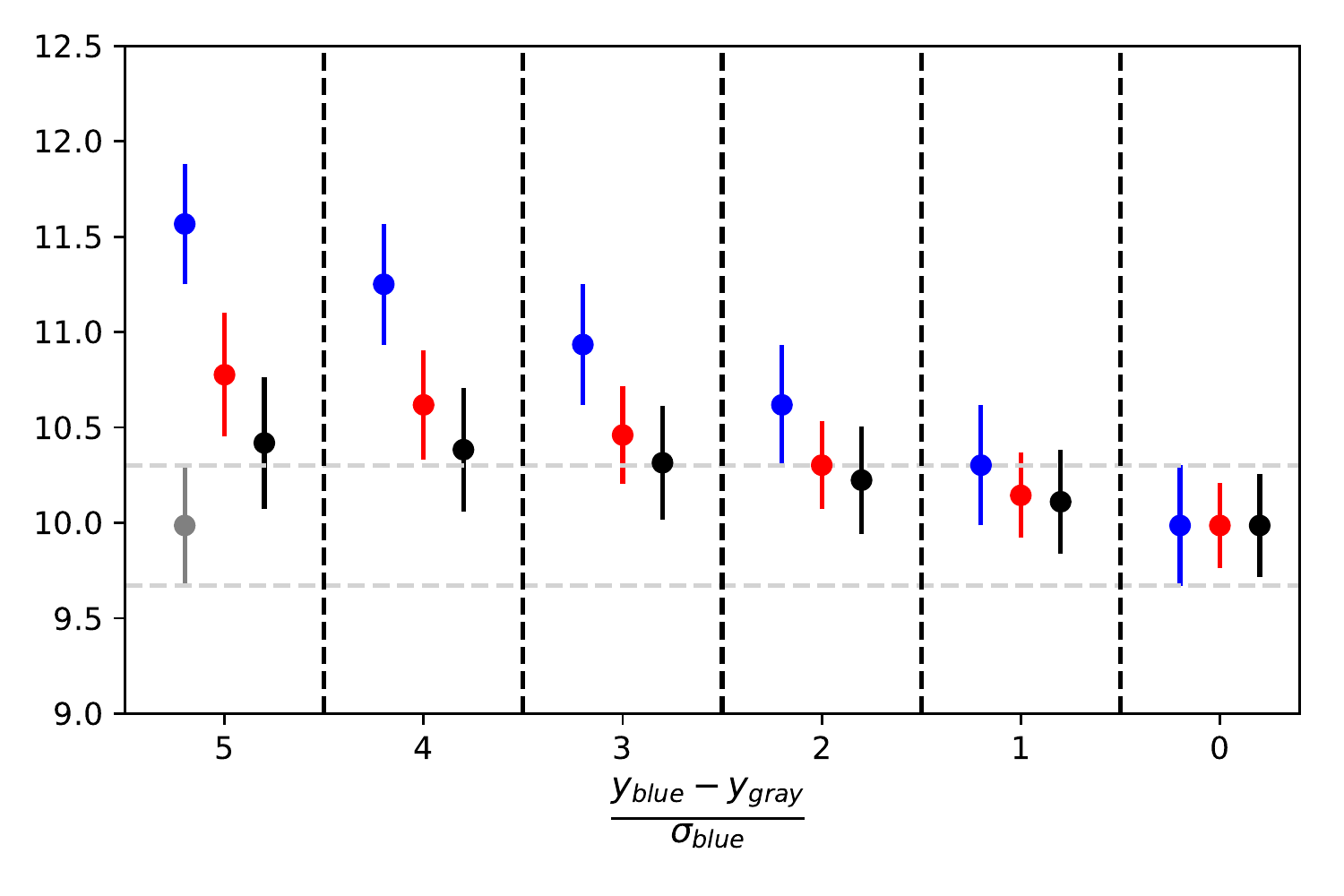}
\caption{The measurement points with small error are shown in blue, the usual averages with the PDG scaling in red, and the hierarchical averages in black. 
The labels at the horizontal axis show by how many $\sigma_{M}$ the blue points deviate from the gray point. 
The gray band represents the ordinary weighted averages of the bulk of measurements in Fig.~\ref{fig:DifferentErros}.}
\label{fig:severalaverages}
\end{centering}
\end{figure}

In Fig.~\ref{fig:severalaverages} we show how the two kind of averages change when we move the central value of the 11th measurement (in blue) 
while leaving the other 10 unchanged.
Just for orientation, the gray band represents the ordinary average (non-hierarchical) of the bulk of measurements with the same error. 
As in Fig.~\ref{fig:DifferentErros}, the red points are the usual PDG-scaled averages, while the black points are the hierarchical averages. 
Clearly, as we approach the bulk the combined error shrinks.

\section{Neutron lifetime}
There is an interesting discrepancy between the two types of experiments measuring the lifetime of the neutron. 
For a state of the art review of both types and more details, see Ref.~\cite{AtomsWietfeldt}.
The first type are beam experiments~\cite{spivak89,beam:1990,Yue:2013qrc},
which measure the number of protons or electrons from decays of cold neutrons in a beam passing through a magnetic or electric trap. 
After the beam has passed the trap, some of the neutrons are deposited in a foil at the end of the beam path. 
The neutron lifetime is proportional to the rate of neutrons deposited and inversely proportional to the rate of decays detected. 

The other type of experiment uses bottles~\cite{Pichlmaier:2010zz,Steyerl:2012zz,Serebrov:2017bzo,Serebrov:2004zf,Ezhov:2014tna,Pattie:2017vsj,Arzumarov2015} 
containing ultra-cold neutrons with a kinetic energy of less than $100$~neV. 
Neutrons with such a low kinetic energy can be confined  due to the effective Fermi potential between neutrons and atomic nuclei in many materials. 
Gravitational forces and magnetic fields can also be used to confine the neutrons within the container. 
The idea is simply to count the number of surviving neutrons after some time and to deduce the lifetime.

\begin{figure}[t]
	\begin{centering}
		\includegraphics[width=0.4\textwidth]{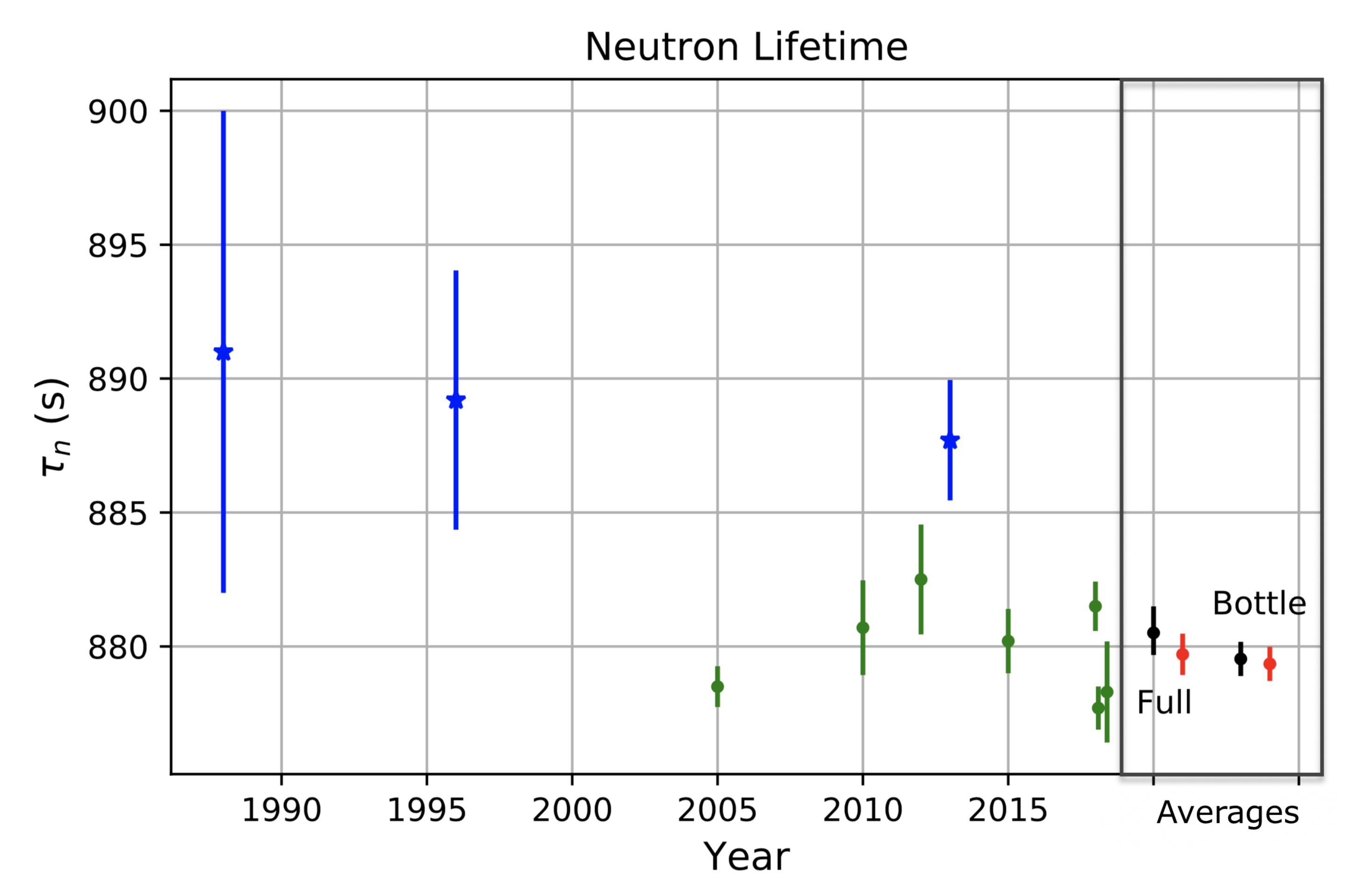}
		\caption{Neutron lifetime measurements. 
		The green points are the results of bottle experiments, and the blue ones of beam experiments. 
		The discrepancy can easily be seen. 
		The black point to the left is the Bayesian average of the full data, while the first red point is the usual average with the PDG scaling. 
		Similarly for the right black and red points but restricted to the bottle results. The PDG scaling for  beam plus bottle experiments  is $S_{PDG}=1.96$, while for bottle only is $S_{PDG}=1.56$.}
		\label{fig:neutronlifetime}
	\end{centering}
\end{figure}
We now apply our method with $\alpha=6$ to the results of these experiments which are shown in Fig.~\ref{fig:neutronlifetime}. 
PDG $\chi^2$ scaling ($S_{PDG}=1.93$), which is shown in red, yields the lifetime $\tau_n=879.71\pm0.78$~s,
while the Bayesian method (black point to the left) gives $\tau^{\rm Bayes}_n=880.51^{+0.98}_{-0.83}$~s.  
We find that our Bayesian hierarchical method increases the central value when the beam experiments are included.
Even when only bottle experiments are considered, 
our method still gives a slightly larger average value $\tau^{\rm Bayes}_n=879.53^{+0.64}_{-0.63}$~s, than the PDG method $\tau_n=879.35\pm0.64$~s where $S_{PDG}=1.56$. 
This is due to the bulk of the bottle experiments that prefer lifetimes longer than 880~s. 
It is important to recall that the tails of the Bayesian hierarchical model do not fall as fast as a Gaussian, 
so that there is still a non-negligible probability for $\tau_n$ to be lower.

\section{Relations to other models}
\label{othermodels}
While this paper was being written, two interesting papers related to our work appeared. 
The first one~\cite{DAgostini:2020vsk} discusses the kaon mass in the context of a skeptical combination of experiments, 
scaling each experimental error independently but correlated.
The second one~\cite{DAgostini:2020pim} studies the discrepancy that arises 
when the PDG scaling is applied to sub-sets of experiments and then to the combination of the sets, {\em vs.\/} (for example) applying it to the whole data at the same time. 
The conclusion is that
\begin{quote}
the $\chi^2/ \nu$ prescription used to enlarge the standard deviation does not hold sufficiency.
\end{quote}
This means that the scaling is not sufficient to properly describe the full probability distribution. 
Our model would have had the same problem had we used the marginalized (over $\tau^2$) distribution of $\mu$.  
This is because the ``correlations" that emerge through $\tau^2$ would be absent. 
But it is clear from Eq.~(\ref{equationtotal}) that if we use the posterior distribution of $\mu$ and $\tau^2$ of a subset 
of experiments as the prior for the remaining subset, then the updated posterior  
will be the same as combining the whole data set simultaneously. 

\begin{figure}[t]
	\begin{centering}
		\includegraphics[width=0.48\textwidth]{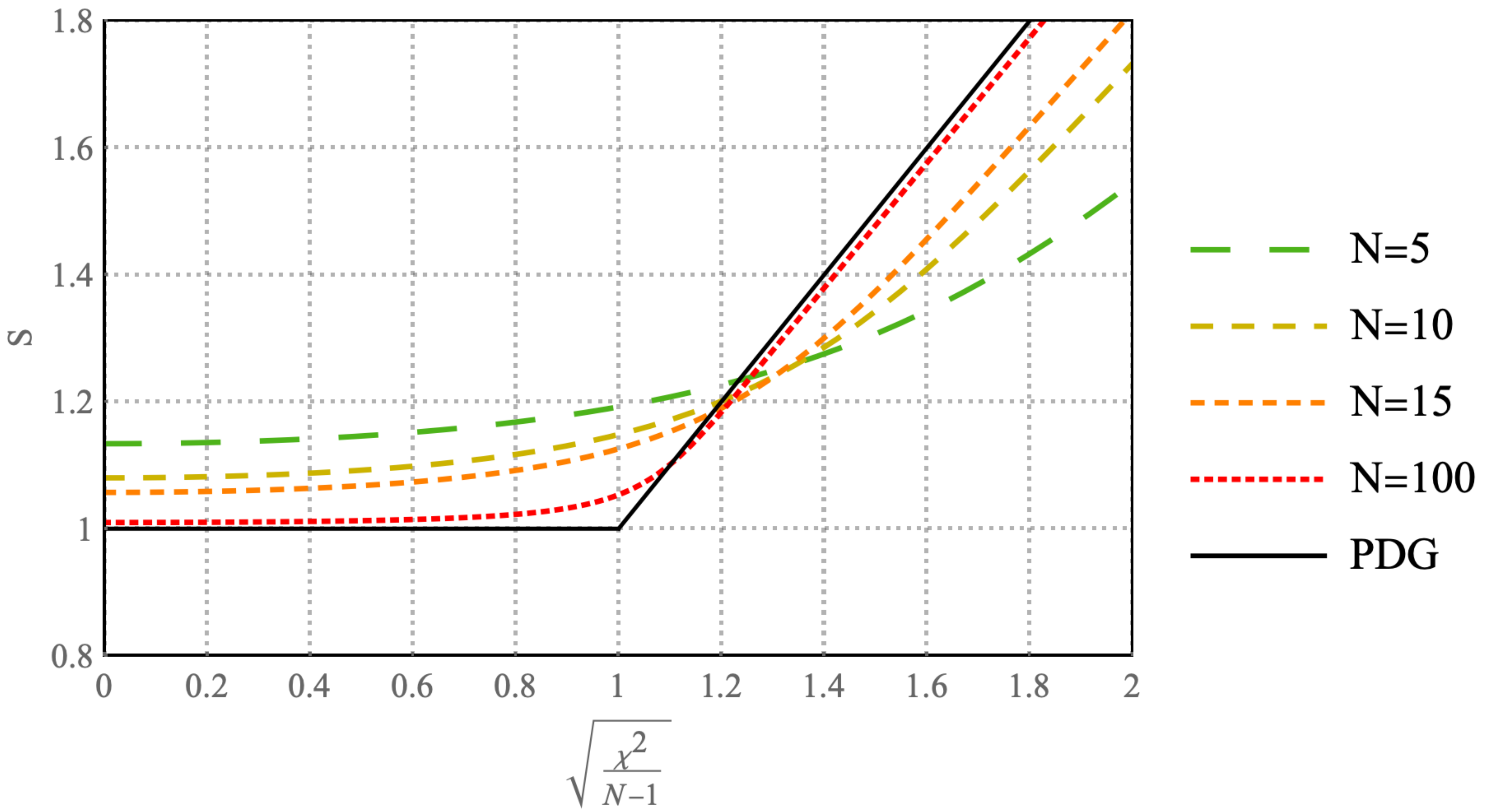}
		\caption{Scaling for $\alpha=6$.}
		\label{fig:scalingalpha6}
	\end{centering}
\end{figure}

Another interesting point made in Ref.~\cite{DAgostini:2020pim} is the fact that the PDG scaling treats any value of $N$ equally,
while for fixed $\chi^2/ N$ the $p$-value decreases with~$N$. 
In other words, since  the probability distribution of the reduced $\chi^2$ function peaks around one as the number of degrees of freedom increases, 
the scaling (given a discrepant value of the reduced $\chi^2$) should be larger when more experiments are included in the average. 
This is not the case for the PDG description, because the scaling only depends on the reduced $\chi^2$ value and not on the number of degrees of freedom.
Now, it is clear from Fig.~\ref{fig:scaleequal} that in the Hierarchical Model with $\alpha$ chosen close to zero this problem would be aggravated,
{\em i.e.}, for any given value of the reduced $\chi^2$, there is more scaling for low $N$.
However, we can use the freedom to choose a value of $\alpha$ to improve on this issue. 
First we demand the variance of the $\tau$ distribution to be finite, which corresponds to $\alpha>6$. 
In Fig.~\ref{fig:scalingalpha6} we show the scaling versus the reduced $\chi^2$ with $\alpha=6+\epsilon$ (where $\epsilon$ is an infinitesimal)  from which one can see 
that for large values of the reduced $\chi^2$ the scaling reduces as $N$ gets smaller. 
This is just the desired effect. 
On the other hand, we still have more scaling for small values of the reduced $\chi^2$. 
This is a natural  consequence of the fact that for a low number of experiments $\tau$ can not be constrained too strongly, which translates into an enlarged error for~$\mu$. 

One can also consider Jeffrey's prior\footnote{In the case of a distribution with several parameters (in our case $\mu$ and $\tau^{2}$), 
Jeffrey's prior is defined as the square root of the determinant of Fisher's information matrix, which in turn is defined as the average
(over $y_{i}$) of the Hessian of the log-likelihood $\mathcal{N}\left(y_{i}|\mu,\tau^{2}+\sigma_{i}^{2}\right)$.}.  
{\em E.g.}, if we specify to the case of uncertainties of equal magnitude, $\sigma_i = \sigma_j =\sigma$, then Jeffrey's prior reduces precisely to Eq.~(\ref{hyperprior}) with $\alpha = 3$. 
This would lead to a plot very similar to the one shown in Fig.~\ref{fig:scalingalpha6}.

\section{Conclusions and outlook}
We proposed a Bayesian hierarchical model as a strategy to compute averages of several uncorrelated experimental measurements,
specifically with the possibility in mind that unaccounted for systematic effects might be present, leading to underestimates of the quoted uncertainties.
We should stress that the point is not that (some part of) the systematic error has been underestimated or assessed too aggressively.
If this is suspected then a strategy should be developed to increase the systematic error component(s),
which would imply --- among other things --- that statistics limited measurements would not be questioned.  
Here, we rather addressed the generic situation in which unknown effects or human errors may be present, 
and which therefore could affect even ostensibly clean determinations. 

We have shown that our methodology resembles the recommendation of the Particle Data Group 
whenever the number of degrees of freedom (data points) is large.
Our approach connects smoothly to cases with fewer degrees of freedom, though.
Another important advantage is that it makes the underlying assumptions in the averaging process transparent.
{\em E.g.}, a large value of the parameter $\alpha$ appearing in our proposed form of the prior, 
implies a strong believe that the experiments do not have an unknown systematic error, while a small value corresponds to a more agnostic point of view. 
Our method can  be extended to experiments with correlated errors, but we leave this generalization for the future.

Due to the additive form, $\sigma_i^2+\tau^2$, of the denominator in the exponential part of the distribution,
our model has the drawback that it tends to penalize experiments with high precision more strongly. 
This relative issue is already seen in the $\tau_n$ example, where the most recent beam measurement which has a larger error 
than most bottle experiments and a higher central value tends to push the combined value up.  
On the other hand, the natural power suppressed tails of the posterior distribution help to mitigate possible strong shifts in the central value. 

We also would like to point out that to apply our method to the PDG, it has to be  studied, discussed and compared with other approaches in more detail, to confirm that it can be used within the PDG framework. 

In closing, we remark that we also envision an application of this model in the context 
of new physics searches within the Standard Model Effective Field Theory (SMEFT) framework~\cite{Buchmuller:1985jz,Grzadkowski:2010es},
in which thousands of {\em a priori\/} independent operator (Wilson) coefficients need to be determined.  
Yet, many of these operators are almost certainly generated at some common energy scale,
and are consequently not entirely independent.
Thus, the idea is to assume that (classes of) the Wilson coefficients are random samples generated at a common ultra-violet energy scale,
lending itself to a hierarchical approach.  
This can be particularly useful when estimating the sensitivity of a hypothetical future experiment to physics beyond the Standard Model.
This is another direction for future work.

\section{Acknowledgments}
We are happy to thank Glen Cowan and Giulio D'Agostini for discussions and comments and Marumi Kado for pointing us to relevant references.
This work was supported by CONACyT (Mexico) project 252167--F, 
and also the German-Mexican research collaboration grant SP 778/4--1 (DFG) and 278017 (CONACyT).

\end{document}